\documentstyle[preprint,aps,psfig]{revtex}
\tighten
\begin{document}
\draft
\preprint{\vbox{Submitted to Physical Review C
                \hfill FSU-SCRI-99-37 \\}}
\title{Quasifree kaon-photoproduction from nuclei \\
       in a relativistic approach}  
\author{L.J. Abu-Raddad and J. Piekarewicz}
\address{Department of Physics and Supercomputer 
         Computations Research Institute, \\ 
         Florida State University, 
         Tallahassee, FL 32306, USA}
\date{\today}
\maketitle
 
\begin{abstract}
 We compute the recoil polarization of the lambda-hyperon and the
photon asymmetry for the quasifree photoproduction of kaons in a
relativistic impulse-approximation approach.  Our motivation for
studying polarization observables is threefold. First, polarization
observables are more effective discriminators of subtle dynamics than
the unpolarized cross section. Second, earlier nonrelativistic
calculations suggest an almost complete insensitivity of polarization
observables to distortions effects. Finally, this insensitivity
entails an enormous simplification in the theoretical
treatment. Indeed, by introducing the notion of a ``bound-nucleon
propagator'' we exploit Feynman's trace techniques to develop
closed-form, analytic expressions for all photoproduction observables.
Moreover, our results indicate that polarization observables are also
insensitive to relativistic effects and to the nuclear target. Yet,
they are sensitive to the model parameters, making them ideal tools
for the study of modifications to the elementary amplitude --- such as
in the production, propagation, and decay of nucleon resonances --- in
the nuclear medium.
\end{abstract}

\narrowtext

\section{Introduction}
\label{sec:intro}
 Impelled by recent experimental advances, there is an increasing
interest in the study of strangeness-production reactions from
nuclei. These reactions form our gate to the relatively unexplored
territory of hypernuclear physics. Moreover, these reactions
constitute the basis for studying novel physical phenomena, such as
the existence of a kaon condensate in the interior of neutron
stars\cite{kn86}. Indeed, the possible formation of the condensate
could be examined indirectly by one of the approved
experiments\cite{chw91} at the Thomas Jefferson National Accelerator
Facility (TJNAF). This experimental approach is reminiscent of the
program carried out at the Los Alamos Meson Physics Facility (LAMPF)
where pion-like modes were studied extensively through the quasifree 
$(\vec{p},\vec{n})$ reaction~\cite{McCl92,Chen93}. These measurements
placed strong constraints on the (pion-like) spin-longitudinal
response and showed conclusively that the long-sought pion-condensed
state does not exist.

The work presented here is a small initial step towards a more
ambitious program that concentrates on relativistic studies of
strangeness in nuclei. Our aim in this paper is the study of the
photoproduction of kaons from nuclei in the quasifree regime. This
investigation helps us in two fronts. First, it sheds light on the
elementary process, $\gamma p \rightarrow K^{+} \Lambda$, by providing
a different physical setting (away from the on-shell point) for
studying the elementary amplitude. Second, it will enable us, in a
future study, to explore modifications to the kaon propagator in the
nuclear medium and to search for those observables most sensitive to
the formation of the condensate. To achieve these goals we focus on
the study of polarization observables. Polarization observables have
been instrumental in the understanding of elusive details about 
subatomic interactions, as they become much more effective
discriminators of subtle physical effects than the traditional
unpolarized cross section. Moreover, quasifree polarization
observables might be one of the cleanest tools for probing the nuclear
dynamics. For example, the reactive content of the process is simple,
as it is dominated by the quasifree production and knockout of a
$\Lambda$-hyperon. Further, free polarization observables provide a
baseline, against which possible medium effects may be inferred. 
Deviations of polarization observables from their free values are 
likely to arise from a modification of the interaction inside the 
nuclear medium or from a change in the response of the target. 
Indeed, relativistic models of nuclear structure predict medium 
modifications to the free observables stemming from an enhanced
lower component of the Dirac spinors in the nuclear
medium~\cite{serwal86}.  Finally, nonrelativistic calculations of the
photoproduction of pseudoscalar mesons suggest that, while distortion
effects provide an overall reduction of the cross section, they do so
without substantially affecting the shape of the
distribution\cite{lwb93,lwbt96,blmw98}. Indeed, these nonrelativistic
calculations show that two important polarization observables --- the
recoil polarization of the ejected baryon and the photon asymmetry ---
are insensitive to distortion effects. Moreover, they seem to be also
independent of the mass of the target nucleus. Thus, quasifree 
polarization observables might represent a fundamental property of 
nuclear matter.

The insensitivity of polarization observables to distortion effects 
is clearly of enormous significance, as one can unravel distortion 
effects from those effects arising from relativity or from the 
large-momentum components in the wavefunction of the bound nucleon.  
Indeed, relativistic plane-wave impulse approximation (RPWIA) 
calculations have been successful in identifying physics not 
present at the nonrelativistic level. For example, relativistic 
effects have been shown to contaminate any attempt to infer color 
transparency from a measurement of the asymmetry in the $(e,e'p)$ 
reaction~\cite{gp94}. Further, the well-known factorization limit 
of nonrelativistic plane-wave calculations has been shown to break 
down due to the presence of negative-energy components in the 
bound-nucleon wavefunction\cite{cdmu98}. Finally, neglecting 
distortions affords the computation of all polarization observables 
in closed form~\cite{gp94} by using the full power of Feynman's trace 
techniques.

We have organized our paper as follows. In Sec.~\ref{sec:formal} we
discuss in detail our relativistic plane-wave formalism, placing
special emphasis on the ``bound-nucleon propagator'' and on the use of
Feynman's trace techniques to evaluate all observables. Our results
are presented in Sec.~\ref{sec:results}, where polarization observables
are computed with two alternative parameterizations of the elementary
amplitude and then compared to results obtained from an on-shell 
nucleon. We offer our conclusions and perspectives for future work 
in Sec.~\ref{sec:concl}.

\section{Formalism}
\label{sec:formal}

\begin{figure}
\bigskip
\centerline{
  \psfig{figure=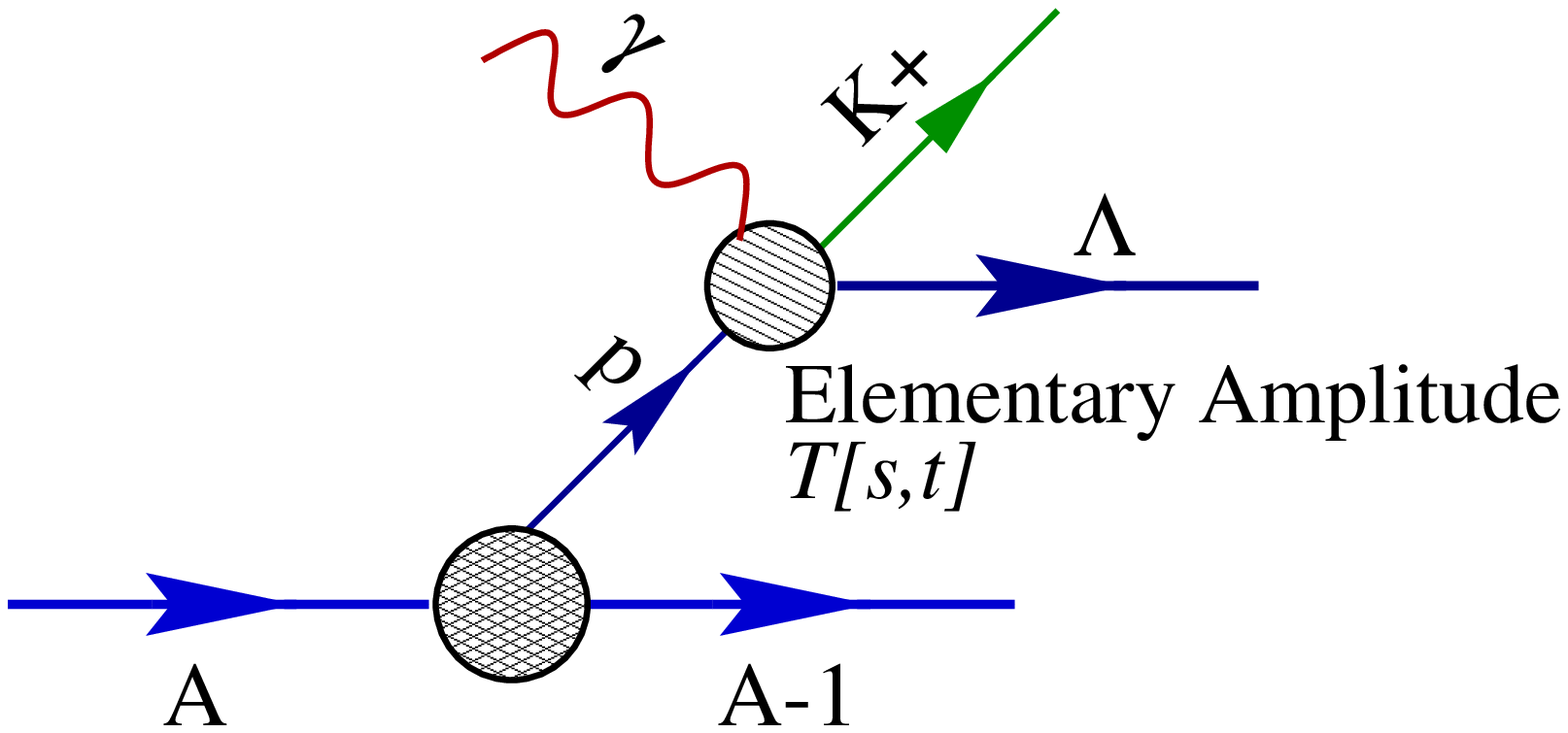,height=2.5in,width=5.5in,angle=0}}
 \vskip 0.1in
 \caption{Representation of the quasifree photoproduction 
	  of a $\Lambda$-hyperon in a plane-wave 
	  impulse-approximation approach.}
 \label{fig1}
\end{figure}

The kinematics for the quasifree production of a $\Lambda$-hyperon 
through the photoproduction reaction $A(\gamma,K^+ \Lambda)B$ is 
constrained by two conditions. First, there is an overall 
energy-momentum conservation:
\begin{equation}
k + p_A = k^{\prime} + p^{\,\prime} + p_B \;.
\end{equation}
Note that $k$ is the four-momentum of the incident photon, while
$k^{\prime}$ and $p^{\prime}$ are the momenta of the produced kaon 
and lambda-hyperon, respectively. Finally, $p_A$($p_B$) represents 
the momentum of the target(residual) nucleus. Moreover, since we are
studying the photoproduction process within the framework of the
impulse approximation (see Fig.~\ref{fig1}) there is a second
kinematical constraint arising from energy-momentum conservation at
the $\gamma N \!\rightarrow\! K^{+} \Lambda$ vertex:
\begin{equation}
k + p = k^{\prime} + p^{\,\prime} \;, 
\end{equation}
where $p$ is the four-momentum of the bound nucleon, whose space
part is known as the missing momentum:
\begin{equation}
{\bf p}_{m}\equiv {\bf p}^{\,\prime}-{\bf q} \;; \quad
                 ({\bf q}\equiv{\bf k}-{\bf k}^{\prime}) \;.
\end{equation}
Thus, as in most semi-inclusive processes --- such as in the $(e,e'p)$ 
reaction --- the quasifree production process becomes sensitive to the 
nucleon momentum distribution. The differential cross section for the 
quasifree process is derived to be:
\begin{equation}
\left(\frac{d^5\sigma (s^{\prime},\varepsilon)}
  {d{\bf k}^{\prime}d\Omega_{{\bf k}^{\prime}}
   d\Omega_{{\bf p}^{\prime}}}\right)_{\rm lab} = 
   \frac{2 \pi}{2 E_\gamma} \mbox{ }
   \frac{|{\bf k}^{\prime}|^2}{(2\pi)^3\,2E_{{\bf k}^{\prime}}}\mbox{ }
   \frac{M_N|{\bf p}^{\prime}|}{(2\pi)^3}\mbox{ } 
   {\big|{\cal M}\big|}^2 \;,
 \label{d5sigma}
\end{equation}
where $s^{\prime}$ is the spin of the emitted $\Lambda$, 
$\varepsilon$ is the polarization of the incident photon, 
$M_N$ is the nucleon mass, and ${\cal M}$ is the transition
matrix element given by:
\begin{equation}
 \big|{\cal M}\big|^2 = \sum_m 
 \Big| 
  \overline{\cal U}({\bf p}^{\prime},s^{\prime})\mbox{ } 
  T(s,t)\mbox{ }{\cal U}_{\alpha,m}({\bf p})
 \Big|^2 \;.
\label{Msquare}
\end{equation}     
Here ${\cal U}({\bf p}^{\prime},s^{\prime})$ is the free
Dirac spinor for the emitted $\Lambda$-hyperon and 
${\cal U}_{\alpha,m}({\bf p})$ is the Fourier transform of 
the relativistic spinor for the bound nucleon ($\alpha$
denotes the collection of all quantum numbers necessary
to specify the single-particle orbital). Note that we 
assume the impulse approximation valid and employ the 
on-shell photoproduction operator $T(s,t)$.

The unpolarized differential expression can be obtained by
summing over the two possible components of the spin of the
$\Lambda$ and averaging over the transverse photon polarization.
That is, 
\begin{equation}
  \left(\frac{d^5\sigma}
  {d{\bf k}^{\prime}d\Omega_{{\bf k}^{\prime}}
   d\Omega_{{\bf p}^{\prime}}}\right)_{\rm lab} = 
   \frac{1}{2}\sum_{s^{\prime},\varepsilon}
   \left(\frac{d^5\sigma (s^{\prime},\varepsilon)}
   {d{\bf k}^{\prime}d\Omega_{{\bf k}^{\prime}}
   d\Omega_{{\bf p}^{\prime}}}\right)_{\rm lab} \;.
\end{equation}
Yet, our prime interest in this work is the calculation of
polarization observables: the recoil $\Lambda$-polarization
(${\cal P}$) and the photon asymmetry ($\Sigma$). The former
is defined as~\cite{wcc90,ndu91}
\begin{equation}
{\cal P} = \sum_{\varepsilon} 
 \left(
  \frac{{d^5\sigma}(\uparrow) - {d^5\sigma}(\downarrow)} 
       {{d^5\sigma}(\uparrow) + {d^5\sigma}(\downarrow)}
 \right)_{\rm lab} \;,
\end{equation}
while the latter by~\cite{lwbt96,blmw98}
\begin{equation}
 {\Sigma} = \sum_{s^{\prime}} 
  \left(
   \frac{{d^5\sigma}(\perp) - {d^5\sigma}(\parallel)} 
        {{d^5\sigma}(\perp) + {d^5\sigma}(\parallel)}
  \right)_{\rm lab} \;.
\end{equation}
In these expressions $\uparrow$ and $\downarrow$ represent the
projection of the spin of the $\Lambda$-hyperon with respect to 
the normal to the scattering plane (${\bf k}^{\prime}\times{\bf k}$),
while $\perp$($\parallel$) represents the out-of-plane(in-plane) 
polarization of the photon.

\subsection{Elementary ($\gamma p \rightarrow K^{+}\Lambda$) Amplitude}
\label{subsec:elemapl}

For the elementary photoproduction amplitude we have used a standard
model-independent parameterization. The parameterization is given in 
terms of four Lorentz- and gauge-invariant 
amplitudes\cite{wcc90,cgln57,bt90,dfls96}
\begin{equation}
 T(\gamma p \rightarrow K^+ \Lambda) = 
 \sum_{i=1}^{4} A_{i}(s,t) M_i\;,
\end{equation}                             
where the invariant matrices have been defined as:
\begin{mathletters}
\begin{eqnarray}
 M_1 &=& - \gamma^5 \rlap/{\varepsilon} \rlap/k \;,   \\ 
 M_2 &=& 2 \gamma^5 [(\varepsilon \cdot p) (k\cdot p^{\prime}) - 
         (\varepsilon\cdot p^{\prime}) (k\cdot p)]\;, \\ 
 M_3 &=& \gamma^5[\rlap/{\varepsilon} ((k\cdot p) - \rlap/k
         (\varepsilon \cdot p)] \; \\ 
 M_4 &=& \gamma^5 [\rlap/{\varepsilon} ((k\cdot p^{\prime}) - 
         \rlap/k (\varepsilon \cdot p^{\prime})] \;.
\end{eqnarray}
\label{M1234}
\end{mathletters}
Although standard, the above parameterization is not unique. Many
other parameterizations --- all of them equivalent on shell --- are 
possible~\cite{raddad99}. It is our hope that the study of quasifree 
polarization observables, together with state-of-the-art measurements, 
will help us elucidate the form of the elementary amplitude.

For the present study, as in all of our earlier photoproduction
studies of pseudoscalar mesons\cite{raddad99,pisabe97,raddad98},
we transform the standard photoproduction amplitude to a more 
suitable form. The resultant form is given by
\begin{equation}
 T(\gamma p \rightarrow K^+ \Lambda) = 
 F^{\alpha \beta}_{T} \sigma_{\alpha \beta} + 
 iF_P\gamma_5 + 
 F^{\alpha}_{A} \gamma _\alpha \gamma_5 \;,
\end{equation}
where tensor, pseudoscalar, and axial-vector amplitudes have been
introduced:
\begin{mathletters}
\begin{eqnarray}
 F^{\alpha \beta}_{T} &=& \frac {1}{2} 
  \varepsilon^{\mu\nu\alpha\beta}\varepsilon_\mu k_\nu A_1(s,t)\;, \\
 F_P &=&-2\,i\Big[(\varepsilon\cdot p) (k\cdot p^{\prime}) 
        -(\varepsilon \cdot p^{\prime}) (k\cdot p)\Big]A_2(s,t)\;, \\
 F^{\alpha}_{A} &=& \Big[(\varepsilon \cdot p) k^\alpha - (k\cdot p) 
                 \varepsilon^\alpha\Big]A_3(s,t) + 
                 \Big[(\varepsilon \cdot p^{\prime})k^\alpha - 
                 (k\cdot p^{\prime})\varepsilon^\alpha\Big]A_4(s,t)\;.
\end{eqnarray}
\end{mathletters}
This form manifests explicitly the Lorentz- and parity-transformation
properties of the various bilinear covariants. Note that in the
presently-adopted parameterization, no scalar nor vector invariant
amplitudes appear. For our calculations we use the hadronic model 
developed by Williams, Ji, and Cotanch\cite{wcc90}. These authors 
impose crossing symmetry in their model to develop phenomenologically
consistent strong-coupling parameterizations which simultaneously
describe the kaon-photoproduction and radiative-capture reactions.

\subsection{Nuclear-Structure Model}
\label{subsec:nsm}

The nuclear-structure model employed here is based on a relativistic
mean-field approximation to the Walecka model\cite{serwal86}. For
spherically symmetric potentials, such as the strong scalar and vector
potentials used here, the eigenstates of the Dirac equation can be
classified according to a generalized angular momentum
$\kappa$\cite{sak73}. These eigenstates can be expressed in a two
component representation; namely,
\begin{equation}
 {\cal U}_{E \kappa m}({\bf x}) = \frac {1}{x} 
 \left[ \begin{array}{c}
   g_{E \kappa}(x) {\cal Y}_{+\kappa \mbox{} m}(\hat{\bf x})  \\
  if_{E \kappa}(x) {\cal Y}_{-\kappa \mbox{} m}(\hat{\bf x})
 \end{array} \right],
\end{equation}
where the spin-angular functions are defined as:
\begin{equation}
 {\cal Y}_{\kappa\mbox{} m}(\hat{\bf x}) \equiv 
 \langle{\hat{\bf x}}|l{\scriptstyle\frac{1}{2}}jm>\;; \quad
 j = |\kappa| - \frac {1}{2} \;; \quad
 l = \cases{   \kappa\;,   & if $\kappa>0$; \cr
            -1-\kappa\;,   & if $\kappa<0$. \cr}
 \label{curlyy}
\end{equation}
The Fourier transform of the relativistic bound-state 
wavefunction --- needed to compute the photoproduction 
cross section [see Eq.~(\ref{Msquare})] --- can now be 
easily evaluated . We obtain,
\begin{equation}
  {\cal U}_{E\kappa m}({\bf p}) \equiv \int d{\bf x} \; 
      e^{-i{\bf p}\cdot{\bf x}}  \; 
      {\cal U}_{E\kappa m}({\bf x}) =
      {4\pi \over p} (-i)^{l}
      \left[
      \begin{array}{c}
       g_{E\kappa}(p) \\
       f_{E\kappa}(p) ({\bf \sigma}\cdot{\hat{\bf p}})
      \end{array}
      \right] {\cal{Y}}_{+\kappa m}(\hat{\bf{p}}) \;,
 \label{uofp}
\end{equation}       
where we have written the Fourier transforms of the radial 
wave functions as
\begin{mathletters}
\begin{eqnarray}
   g_{E\kappa}(p) &=& 
    \int_{0}^{\infty} dx \,g_{E\kappa}(x)  
    \hat{\jmath}{\hbox{\lower 3pt\hbox{$_l$}}}(px) \;, \\
   f_{E\kappa}(p) &=& ({\rm sgn}\kappa) 
    \int_{0}^{\infty} dx \,f_{E\kappa}(x)
    \hat{\jmath}{\hbox{\lower 3pt\hbox{$_{l'}$}}}(px) \;.
\end{eqnarray}
\label{gfp}
\end{mathletters}
Note that in the above expression we have introduced
the Riccati-Bessel function in terms of the spherical
Bessel function~\cite{Taylor72}:
$\hat{\jmath}{\hbox{\lower 3pt\hbox{$_{l}$}}}(z)=
zj{\hbox{\lower 3pt\hbox{$_{l}$}}}(z)$ and that
$l'$ is the orbital angular momentum corresponding
to $-\kappa$ [see Eq.~(\ref{curlyy})].

\subsection{Closed-form Expression for the Photoproduction Amplitude}
\label{subsec:closedform}

Having introduced all relevant quantities, we are now in a position
to evaluate the (square of the) photoproduction amplitude 
[Eq.~(\ref{Msquare})]. As mentioned earlier, polarization observables 
are computed within the framework of a relativistic PWIA. That this is, 
indeed, an excellent approximation seems to be justified by the 
nonrelativistic analysis presented in Refs.~\cite{lwb93,lwbt96,blmw98}.
Without distortions, the evaluation of the $\Lambda$ propagator is now
standard due to an algebraic ``trick'' that appears to be used for the 
first time by Casimir~\cite{Pais86,g87}:
\begin{equation}
 S(p^{\prime}) \equiv \sum_{s^{\prime}}
            {\cal U } ({\bf p}^{\prime},s^{\prime}) \,
  \overline{{\cal U }}({\bf p}^{\prime},s^{\prime}) =
  \frac{\rlap/p^{\prime}+M_{\Lambda}}{2M_{\Lambda}} \;; \quad
  \left(p^{\prime\,0}\equiv E_{\Lambda}({\bf p}^{\prime})=
   \sqrt{{\bf p}^{\prime\,2}+M_{\Lambda}^2}\right) \;.
\label{sfree}
\end{equation}
Subsequently others --- Feynman being apparently the first one ---
used this trick to reduce the ``complex'' computation of covariant
matrix elements to a simple and elegant evaluation of traces of
Dirac $\gamma$-matrices. These trace-techniques have been used here to
compute free polarization observables (note that free polarization
observables will serve as the baseline for comparison against 
bound-nucleon calculations). In principle, one does not expect that 
these useful trace-techniques will generalize once the nucleon goes 
off its mass shell. Yet, simple algebraic manipulations --- first 
performed to our knowledge by Gardner and Piekarewicz~\cite{gp94} --- 
show that a trick similar to that of Casimir holds even for bound 
spinors. Indeed, the validity of their result rests on the following 
simple identity:
\begin{equation}
  \sum_{m}
   {\cal{Y}}_{+\kappa m}(\hat{\bf{p}})  
   {\cal{Y}}^{\star}_{\pm\kappa m}(\hat{\bf{p}})  
   = \pm\frac{2j+1}{8\pi}
     \cases{1 \cr 
           {\bf \sigma}\cdot\hat{\bf p} \cr} \;,
\end{equation}
which enables one to introduce the notion of a 
``bound-state propagator''. That is,
\begin{eqnarray}
  S_{\alpha}({\bf p})  
  &\equiv& {1 \over 2j+1} \sum_{m} 
               {\cal U}_{\alpha,m}({\bf p}) \,
      \overline{\cal U}_{\alpha,m}({\bf p}) \nonumber \\
  &=& \left({2\pi \over p^{2}}\right)
      \left( 
       \begin{array}{cc}
         g^{2}_{\alpha}(p) & 
        -g_{\alpha}(p)f_{\alpha}(p){\bf \sigma}\cdot{\hat{\bf p}} \\
        +g_{\alpha}(p)f_{\alpha}(p){\bf \sigma}\cdot{\hat{\bf p}} &
        -f^{2}_{\alpha}(p) 
       \end{array}
      \right) \nonumber \\
  &=& ({\rlap/{p}}_{\alpha} + M_{\alpha}) \;, \quad
      \Big(\alpha=\{E,\kappa\}\Big) \;.
 \label{salpha}
\end{eqnarray}
Note that we have defined the above mass-, energy-, and momentum-like
quantities as
\begin{mathletters}
\begin{eqnarray}
  M_{\alpha} &=& \left({\pi \over p^{2}}\right)
                  \Big[g_{\alpha}^{2}(p) -
                       f_{\alpha}^{2}(p)\Big] \;, \\
  E_{\alpha} &=& \left({\pi \over p^{2}}\right)
                  \Big[g_{\alpha}^{2}(p) +
                       f_{\alpha}^{2}(p)\Big] \;,
 \label{epm} \\
  {\bf p}_{\alpha} &=& \left({\pi \over p^{2}}\right)
                   \Big[2 g_{\alpha}(p) 
                          f_{\alpha}(p)\hat{\bf p} 
                   \Big]  \;,
\end{eqnarray}
\end{mathletters}
which satisfy the ``on-shell relation''
\begin{equation}
  p_{\alpha}^{2}=E_{\alpha}^{2}-{\bf p}_{\alpha}^{2}
                =M_{\alpha}^{2} \;.
 \label{onshell}
\end{equation}
The evident similarity in structure between the free and bound
propagators [Eqs.~(\ref{sfree}) and (\ref{salpha})] results in an
enormous simplification; we can now employ the powerful trace
techniques developed by Feynman to evaluate all polarization
observables --- irrespective if the nucleon is free or bound to a
nucleus.  It is important to note, however, that this enormous
simplification would have been lost if distortion effects would have
been incorporated in the propagation of the emitted $\Lambda$-hyperon
or $K^{+}$-meson.

To provide a feeling for the enormous simplification entailed by the
above trick, we derive now an example assuming, for mere simplicity, 
that the photoproduction amplitude contains only the tensor term 
[$M_{1}$ in Eq.~(\ref{M1234})]. For this simplified case the square 
of the unpolarized photoproduction matrix element [Eq.~(\ref{Msquare})] 
becomes proportional to:
\begin{eqnarray}
 \big|{\cal M}\big|^2 &\rightarrow& |A_1|^2 \, 
 \left(-\frac{1}{2}g_{\mu\nu}\right)\,
 T{\hbox{\lower 2pt\hbox{$r$}}}
 \Big[\gamma^{5}\gamma^{\mu}\rlap/k
        \left(\rlap/p_{\alpha}+M_{\alpha}\right)
        \gamma^{5}\gamma^{\nu}\rlap/k
        \left(\rlap/p^{\prime}+M_{\Lambda}\right)
      \Big] \nonumber \\
 &=& \frac{1}{2}\,|A_1|^2 \left[ 
        T{\hbox{\lower 2pt\hbox{$r$}}}
        \Big(\gamma^{\mu}\rlap/k\rlap/p_{\alpha}
             \gamma_{\mu}\rlap/k\rlap/p^{\prime}
        \Big) - M_{\alpha}M_{\Lambda}
        T{\hbox{\lower 2pt\hbox{$r$}}}
        \Big(\gamma^{\mu}\rlap/k
             \gamma_{\mu}\rlap/k 
        \Big) \right]\nonumber \\
 &=& 8\,|A_1|^2  
       (k \cdot p_{\alpha})(k\cdot p^{\prime}) \;.
\end{eqnarray}    
This result is, indeed, simple and illuminating. Although including
the full complexity of the elementary amplitude requires the
evaluation of many such terms (not all of them independent) the
evaluation of any one of those terms is not much more complicated than
the one presented above. Yet, to automate this straightforward but
lengthy procedure, we rely on the {\it FeynCalc 1.0}\cite{mh92}
package with {\it Mathematica 2.0} to calculate all traces involving
$\gamma$-matrices. The output from these symbolic manipulations was
then fed into a FORTRAN code to obtain the final numerical values for
all different polarization observables.

\section{Results and Discussion}
\label{sec:results}
\begin{figure}
\bigskip
\centerline{
  \psfig{figure=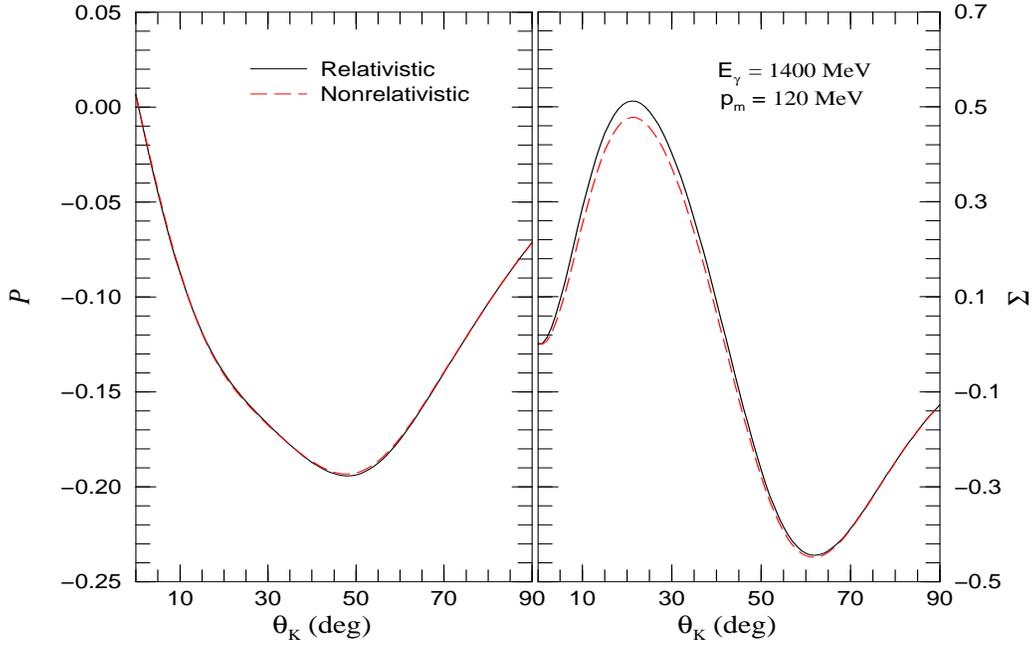,height=3.5in,width=5.5in,angle=-90}}
 \vskip 0.1in
 \caption{Comparison between relativistic and nonrelativistic
	  calculations of the recoil polarization of the 
	  $\Lambda$-hyperon (${\cal P}$) and the photon 
	  asymmetry ($\Sigma$) as a function of the kaon 
	  scattering angle for the knockout of a proton 
	  from the $p^{3/2}$ orbital in ${}^{12}$C.} 
 \label{fig2}
\end{figure}

We start the discussion of our results by examining the role of 
the relativistic dynamics on the polarization observables. On
Fig.~\ref{fig2} we display the recoil polarization (${\cal P}$) 
of the $\Lambda-$hyperon and the photon asymmetry ($\Sigma$) as 
a function of the kaon scattering angle for the knockout of a 
proton from the $p^{3/2}$ orbital in ${}^{12}$C. The polarization 
observables were evaluated at a photon energy of 
$E_{\gamma}\!=\!1400$~MeV and at a missing momentum of 
$p_m\!=\!120$~MeV (this value is close to the maximum in the 
momentum distribution of the $p^{3/2}$ orbital; see
Fig.~\ref{fig6}). Note that in this figure --- and all throughout 
this paper --- we compute all observables in the laboratory system
using the quasifree condition:
$\omega\!=\!\sqrt{q^2+M_{\Lambda}^{2}}-M_{N}$. The insensitivity 
of our results to the relativistic dynamics is evident. Indeed, 
for the case of the recoil polarization the two curves cannot 
even be resolved in the figure. We have also examined these 
effects on the unpolarized cross section and found them 
insignificant as well. Note that our ``nonrelativistic'' results 
were obtained by adopting the free-space relation in the 
determination of the lower-components of the bound-state 
wavefunction. This represents our best attempt at reproducing 
nonrelativistic calculations, which employ free, on-shell spinors 
to effect the nonrelativistic reduction of the elementary 
amplitude.

\begin{figure}
\bigskip
\centerline{
  \psfig{figure=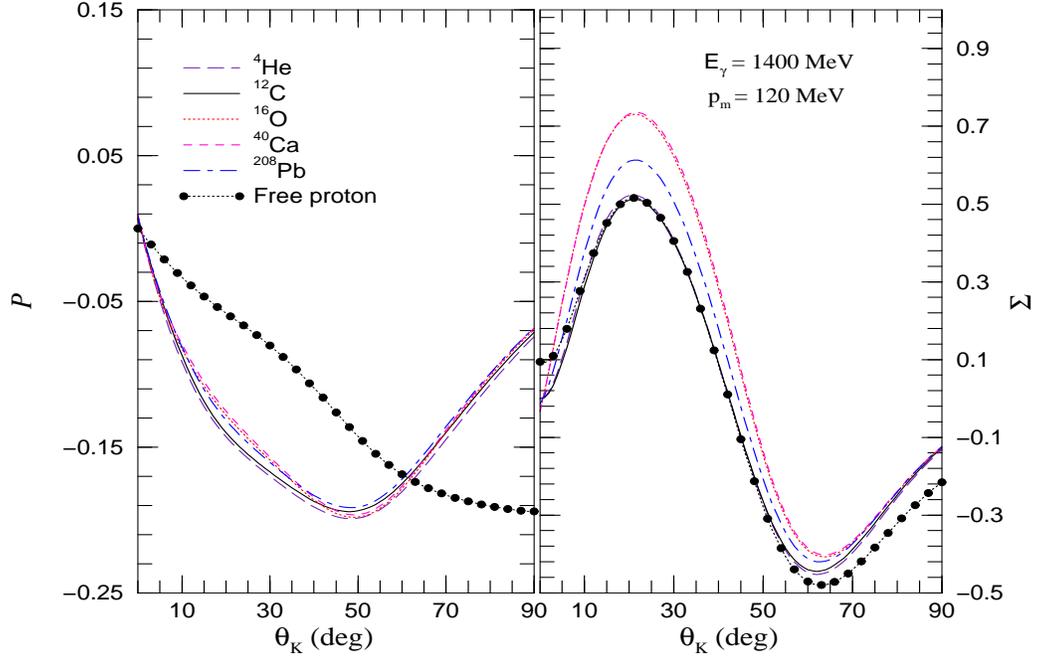,height=3.5in,width=5.5in,angle=-90}}
 \vskip 0.1in
 \caption{The recoil polarization of the $\Lambda$-hyperon 
          (${\cal P}$) and the photon asymmetry ($\Sigma$)
	  as a function of the kaon scattering angle for the
	  knockout of a valence proton from a variety of nuclei.
	  The photoproduction from a free proton is depicted
	  with the filled circles.}
 \label{fig3}
\end{figure}

Next we examine the nuclear dependence of the polarization
observables. Fig.~\ref{fig3} displays the recoil polarization and the
photon asymmetry for the knockout of a valence proton for a variety of
nuclei, ranging from ${}^{4}$He all the way to ${}^{208}$Pb. That is,
we have computed the knockout from the $1s^{1/2}$ orbital of
${}^{4}$He, the $1p^{3/2}$ orbital of ${}^{12}$C, the $1p^{1/2}$
orbital of ${}^{16}$O, the $1d^{3/2}$ orbital of ${}^{40}$Ca, and the
$3s^{1/2}$ orbital of ${}^{208}$Pb. We have included also polarization
observables from a single proton to establish a baseline for
comparison against our bound--nucleon calculations. The insensitivity
of the recoil polarization (${\cal P}$) to the nuclear target is
striking, indeed. As soon as the quasifree process takes place from a
proton bound to a ``lump'' of nuclear matter, the recoil polarization
becomes insensitive to the fine details of the lump.  Moreover, the
deviations from the free value (shown with the filled circles) are
significant. This indicates important modifications to the elementary
amplitude in the nuclear medium and suggests that the recoil
polarization of the $\Lambda$-hyperon might represent a fundamental
property of nuclear matter. For the photon asymmetry ($\Sigma$) the
effect, although still significant, does not to seem to be as
impressive as for the case of the recoil polarization. For example,
$\Sigma$ is no longer independent of the nuclear target. While
${}^{4}$He and ${}^{12}$C are almost identical to the free case,
${}^{16}$O and ${}^{40}$Ca --- while practically indistinguishable
from each other --- depart significantly from the free
value. Moreover, the photon asymmetry does not seem to saturate, as
evince by the ${}^{208}$Pb results. On the other hand, the shape of
the photon asymmetry is preserved in going from the free nucleon 
value all the way to ${}^{208}$Pb.

Having established the independence of polarization observables 
to the nuclear target and to relativistic effects --- particularly 
for the case of the recoil polarization --- we are now in a good
position to discuss the sensitivity of these observables to the
elementary amplitude (note that the insensitivity of polarization 
observables to final-state interaction has been shown convincingly 
in Ref.~\cite{blmw98}). In the crossing-symmetric model of 
Ref.~\cite{wcc90} two equally good fits to kaon-photoproduction 
and radiative-capture data were provided. These are the two sets, 
labeled ``C1'' and ``C2'', that we adopt here. We note that other 
choices --- perhaps more sophisticated and up to date --- can be 
easily incorporated into our calculation.

\begin{figure}
\bigskip
\centerline{
  \psfig{figure=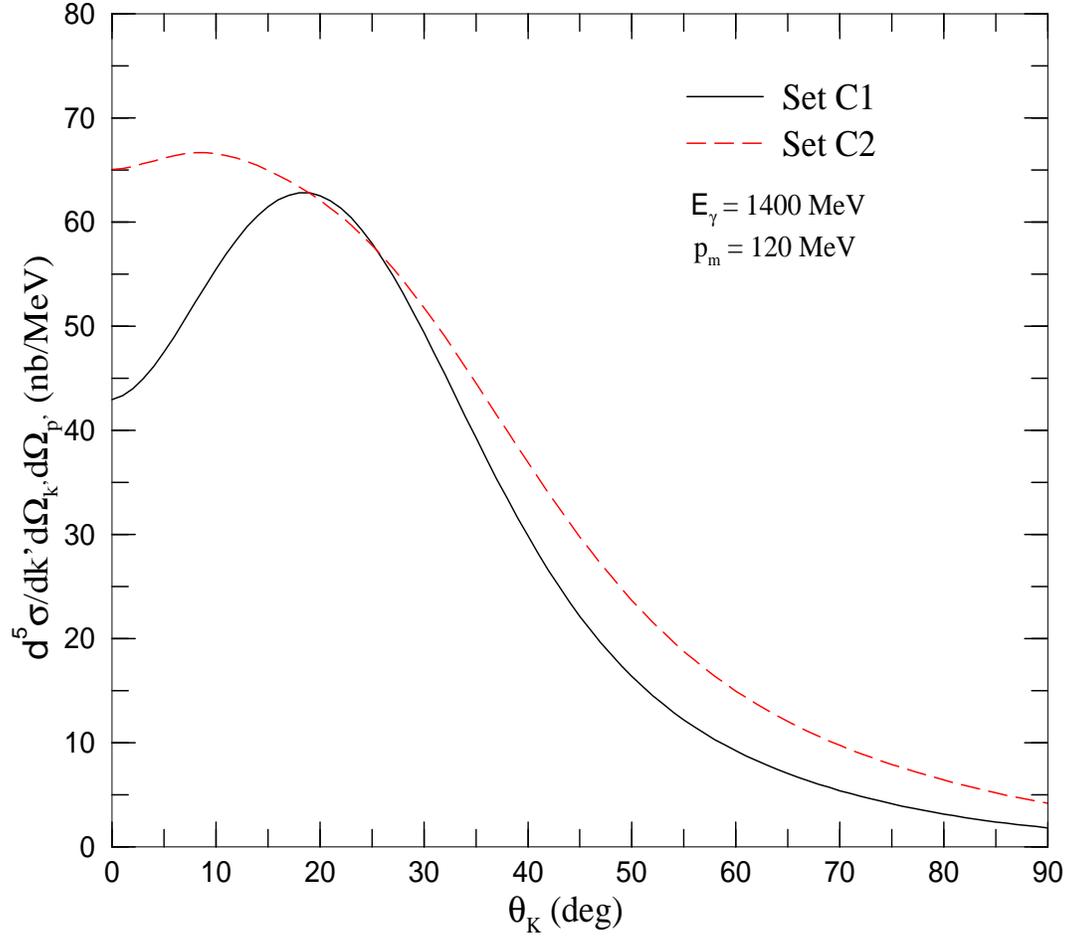,height=5in,width=5.5in,angle=-90}}
 \vskip 0.1in
 \caption{The differential cross section as a function of the kaon
          scattering angle for the knockout of a proton from the 
	  $p^{3/2}$ orbital in ${}^{12}$C using two different sets
	  for the elementary amplitude: C1 and C2.}
 \label{fig4}
\end{figure}

\begin{figure}
\bigskip
\centerline{
  \psfig{figure=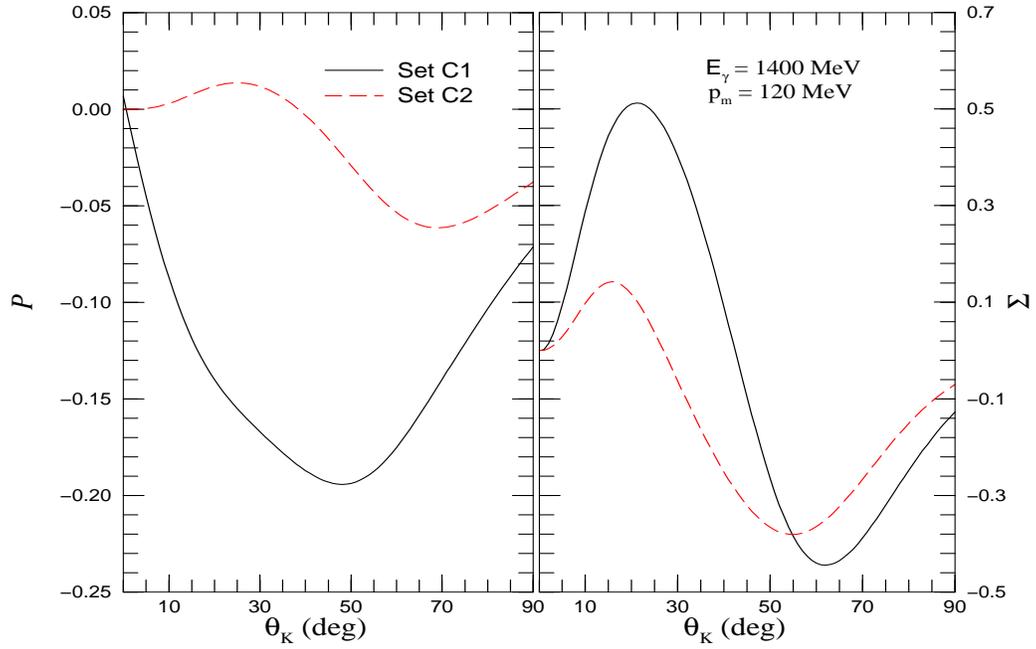,height=3.5in,width=5.5in,angle=-90}}
 \vskip 0.1in
 \caption{The recoil polarization of the $\Lambda$-hyperon
	  (${\cal P}$) and the photon asymmetry ($\Sigma$) 
	  as a function of the kaon scattering angle
	  for the knockout of a proton from the 
	  $p^{3/2}$ orbital in ${}^{12}$C using two 
          parameterizations of the elementary amplitude: 
	  C1 and C2.}
 \label{fig5}
\end{figure}

\begin{figure}
\bigskip
\centerline{
  \psfig{figure=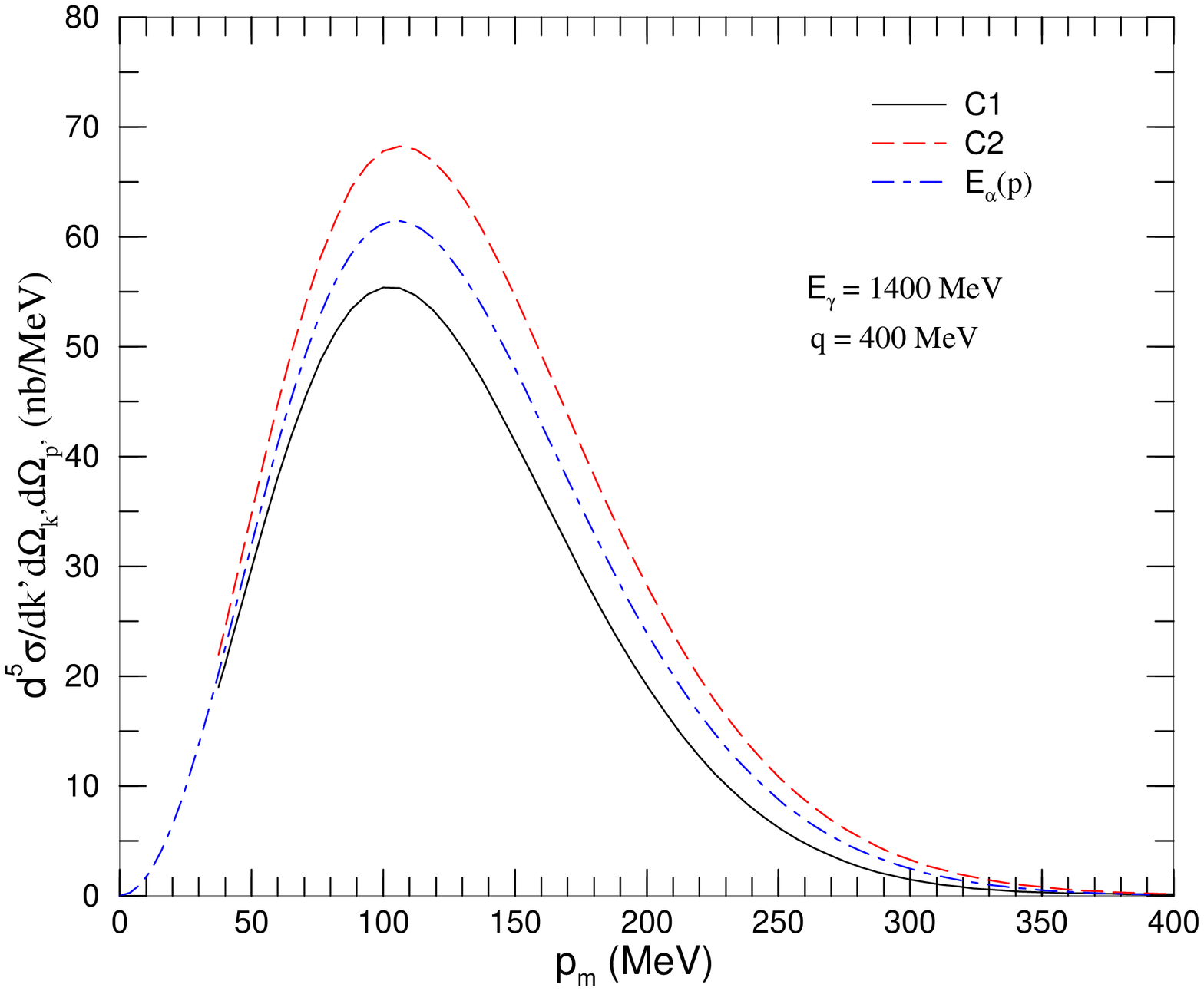,height=5in,width=5.5in,angle=-90}}
 \vskip 0.1in
 \caption{The differential cross section as a function of the missing
          momentum for the knockout of a proton from the $p^{3/2}$ 
	  orbital of ${}^{12}$C using two parameterizations of the 
	  elementary amplitude: C1 and C2. The figure includes also 
	  the $E_\alpha$ parameter (up to an arbitrary scale) which is
	  proportional to the momentum distribution of the
          bound-proton wavefunction (see text for details).}
 \label{fig6}
\end{figure}

We display in Fig.~\ref{fig4} the differential cross section as a
function of the kaon scattering angle for the knockout of a proton
from the $p^{3/2}$ orbital in ${}^{12}$C. Again, the photon incident
energy and the missing momentum have been fixed at $1400$~MeV and
$120$~MeV, respectively. Although there are noticeable differences
between the two sets at small angles, these differences essentially
disappear past $\theta_{K}\simeq 20^{\circ}$. Much more significant,
however, are the differences between the two sets for the case of the
polarization observables displayed in Fig.~\ref{fig5}. The added
sensitivity to the choice of amplitude exhibited by the polarization
observables should not come as a surprise; unraveling subtle details
about the nuclear dynamics is the hallmark of polarization
observables.  We have compared our results in
Figs.~\ref{fig4}-\ref{fig5} to the nonrelativistic results of
Ref.~\cite{blmw98} and have found nice agreement between them, 
particularly in the case of the set C1. Both calculations predict 
similar behavior for all the observables and differ only in the 
fine details. Knowing that relativity plays no role in the 
calculation --- and knowing that the polarization observables 
are very sensitive to the choice of amplitude --- the differences 
between the two sets of calculations must lie solely in the choice 
of elementary amplitude.
	
Finally, we display in Fig.~\ref{fig6} the cross section as a function
of missing momentum for the $p^{3/2}$ orbital in ${}^{12}$C using a
different kinematical setting. Here we have kept the photon incident
energy fixed at $1400$~MeV but have set the momentum transfer $q$ at
$400$~MeV. To a large extent the cross sections represents --- up to
an overall normalization factor --- the momentum distribution of the
$p^{3/2}$ orbital. Indeed, the peak in the cross section is located at
$p_{m}\!\approx\!110$~MeV, which is also the position of the maximum
in the momentum distribution. Moreover, to further appreciate the
similarities between the two we have included the energy-like
parameter $E_\alpha$, up to an arbitrary scale. As seen from
Eq.~(\ref{epm}), $E_\alpha$ is directly proportional to the momentum
distribution of the bound-proton wavefunction. The similarities
between the cross section and the momentum distribution are
indisputable. Note that the cross section dies out for $p_{\rm
m}\!>\!250$~MeV. This region of high-momentum components is sensitive
to short-range correlations, which are beyond the scope of our simple
mean-field description. Thus, the tail of the photoproduction cross
section can be used to test more sophisticated models of nuclear
structure.

\section{Conclusions}
\label{sec:concl}
We have computed polarization observables --- the recoil polarization
of the $\Lambda$-hyperon and the photon asymmetry --- for the
quasifree $K^{+}$ photoproduction reaction from nuclei. Due to the
insensitivity of polarization observables to distortion effects, a
relativistic plane-wave impulse approximation was developed. For the
elementary amplitude we used the relativistic crossing-symmetric model
developed by Williams, Ji, and Cotanch in Ref.~\cite{wcc90}, while for
the nuclear structure we employed a relativistic mean-field
approximation to the Walecka model~\cite{serwal86}. In this manner the
quasifree amplitude was evaluated without recourse to a nonrelativistic 
reduction, as the full relativistic structure of the amplitude was
maintained.  

By assuming the validity of the relativistic plane-wave impulse
approximation an enormous simplification ensued: by introducing the 
notion of a bound-state propagator --- as was done for the first
time by Gardner and Piekarewicz in Ref.~\cite{gp94} --- the
mathematical structure of all quasifree observables was casted in a 
manner analogous to that of the elementary process. Thus, we brought 
the full power of Feynman's trace techniques to bear into the problem.
We stress that the relativistic formalism presented here can be
applied with minor modifications to most quasifree knockout studies,
at least in the plane-wave limit.

In addition of being insensitive to distortions effects, we found
polarization observables almost independent of the target nucleus and
insensitive to relativistic effects. The only effect that polarization
observables appear to be sensitive to is the elementary amplitude.  As
free polarization observables provide a baseline against which
possible medium effects may be inferred, we conclude that quasifree
polarization observables might be one of the cleanest tools for
probing modifications to the elementary amplitude in the nuclear
medium. Deviations from their free values are likely to stem from a
modification of the elementary interaction inside the nuclear medium
due, for example, to a change in resonance parameters.  Indeed, for
the kinematics adopted in this work ($E_{\gamma}\!=\!1.4$~GeV or
$\sqrt{s}\!\approx\!1.9$~GeV) one should be very sensitive to the
formation, propagation, and decay of the $P_{13}(1900)$ and
$F_{17}(1990)$ $N^{\star}-$resonances~\cite{PDG98}. The meson
photoproduction (and electroproduction) programs at various
experimental facilities --- such as TJNAF, NIKHEF, and MAMI --- 
should shed light into the physics of this interesting and 
fundamental problem.

\acknowledgments
This work was supported in part by the United States Department of
Energy under Contracts Nos. DE-FC05-85ER250000 and DE-FG05-92ER40750.

\vfill\eject
\vfill\eject


\begin{references}
\bibitem{kn86}      D.B. Kaplan and A.E. Nelson, Phys. Lett. B {\bf
                    175}, (1986), 57; B {\bf 179}, (1986), 409(E).   
\bibitem{chw91}     C. Hyde-Wright,{\it Quasifree Strangeness
	            Production in Nuclei}, Hall B, Experiment Number {\bf
	            E-91-014}.
\bibitem{McCl92}    J.B. McClelland {\it et al.,} Phys. Rev. Lett.
                    {\bf 69}, 582 (1992).
\bibitem{Chen93}    X.Y. Chen {\it et al.,} Phys. Rev. C~{\bf 47},
                    2159 (1993).
\bibitem{serwal86}  J.D.~Walecka, Ann. of Phys. (N.Y.) {\bf 83}, 491 (1974);
                    B.D. Serot and J.D. Walecka, Adv. in Nucl. Phys. 
                    {\bf 16}, J.W. Negele and E. Vogt, eds. 
                    Plenum, N.Y. (1986).
\bibitem{lwb93}     X. Li(F.X. Lee), L.E. Wright, and C. Bennhold,
		    Phys. Rev. C~{\bf 48}, 816 (1993).
\bibitem{lwbt96}    F.X. Lee, L.E. Wright, C. Bennhold, and
		    L. Tiator, Nucl.~Phys.~{\bf A603}, 345 (1996).
\bibitem{blmw98}    C. Bennhold, F.X. Lee, T. Mart, and L.E. Wright,
		    Nucl.~Phys.~{\bf A639}, 227 (1998). 
\bibitem{gp94}	    S. Gardner, and J. Piekarewicz,
		    Phys. Rev. C~{\bf 50}, 2822 (1994).
\bibitem{cdmu98}    J.A. Caballero, T.W. Donnelly, E. Moya de Guerra,
		    and J.M. Udias,  Nucl.~Phys.~{\bf A632}, 323
		    (1998).
\bibitem{wcc90}     R. Williams, C.R. Ji, and S. R. Cotanch,
		    Phys. Rev. D~{\bf 41}, 1449 (1990).
\bibitem{ndu91}     A. Nagl, V. Devanathan, and H. \"Uberall,
		    {\it Nuclear Pion Photoproduction} Springer Tracts in
		    Modern Physics {\bf 120}, Springer-Verlag, Berlin
		    Heidelberg, Germany (1991). 
\bibitem{cgln57}    G.F. Chew, M.L. Goldberger, F.E. Low, and Y. Nambu,
                    Phys. Rev. {\bf 106}, 1345 (1957).
\bibitem{bt90}      C. Bennhold and H. Tanabe,
                    Phys. Lett. B~{\bf 243}, 13 (1990);
                    Nucl.~Phys.~{\bf A530}, 625 (1991).
\bibitem{dfls96}    J.C. David, C. Fayard, G.H. Lamot, and B. Saghai,
		    Phys. Rev. C~{\bf 53}, 2613 (1996).
\bibitem{raddad99}  L.J. Abu-Raddad, J. Piekarewicz, A. J. Sarty, and 
	            R.A. Rego, {\tt nucl-th/9812061}.
\bibitem{pisabe97}  J. Piekarewicz, A. J. Sarty, and M. Benmerrouche,
		    Phys. Rev. C~{\bf 55}, 2571 (1997).
\bibitem{raddad98}  L.J. Abu-Raddad, J. Piekarewicz, A. J. Sarty, and 
	            M. Benmerrouche, Phys. Rev. C~{\bf 57}, 2053 (1998).
\bibitem{sak73}     J.J. Sakurai, {\it Advanced Quantum Mechanics},
		    Addison-Wesley Publishing Company, 1973. 
\bibitem{Taylor72}  John R. Taylor, {\it Scattering Theory: The
	            Quantum Theory of Nonrelativistic Collisions}
		    (John Wiley \& Sons, Inc., New York, 1972).	
\bibitem{Pais86}    A. Pais, {\it Inward Bound}, p. 375 
                    (Oxford, New York, 1986).
\bibitem{g87}	    D. Griffiths,{\it Introduction to Elementary
		    Particles} (John Wiley \& Sons, Inc., Singapore,
		    1987).
\bibitem{mh92}      R. Mertig and A. Hubland, {\it Guide to FeynCalc
                    1.0}, downloaded from the internet, 1992.  
\bibitem{PDG98}     C. Caso {\it et al.,} (Particle Data Group) 
	            The European Physical Journal C3 (1998) 1. 
\end{references}
\end{document}